\documentclass[%
 reprint,
 amsmath,amssymb,
 aps,
]{revtex4-2}

\usepackage{hyperref}
\usepackage{graphicx}

\usepackage{amsmath}
\usepackage{amssymb}
\usepackage{dcolumn}
\usepackage{bm}
\usepackage[thinc]{esdiff}

\begin{document}

\preprint{APS/123-QED}

\title{A Braess' paradox analog in physical networks of optimal exploration}

\author{Georgios Gounaris}
\author{Eleni Katifori}%
\affiliation{Department of Physics and Astronomy, University of Pennsylvania, Philadelphia, PA.}

\date{\today}

\begin{abstract}

In the stochastic exploration of geometrically embedded graphs, intuition suggests that providing a shortcut between a pair of nodes reduces the mean first passage time of the entire graph. Counterintuitively, we find a Braess' paradox analog. For regular diffusion, shortcuts can worsen the overall search efficiency of the network, although they bridge topologically distant nodes. We propose an optimization scheme under which each edge adapts its conductivity to minimize the graph's search time. The optimization reveals a relationship between the structure and diffusion exponent and a crossover from dense to sparse graphs as the exponent increases.

\end{abstract}

\maketitle

Time is a limiting factor for a plethora of physical networks that rely on diffusion as a mechanism of transport and search. The diffusive exploration can describe Euclidean space trajectories in various length scales, e.g. foraging animals searching for food or molecules searching for a binding target \cite{Intermittent_search}. Diffusion is also used to describe the stochastic transitions between the numerous meta-stable states in an energy landscape \cite{energy_landscapes}. In all of these different systems, the average timescale in which a random walker will encounter a target for the first time is given by the mean first passage time (MFPT) \cite{redner_2001}.

The diffusive exploration of complex networks is inherently linked to the structural properties of the graph  \cite{bassolas,Morozov}.  For instance, a random walk  on a regular lattice becomes transient once the lattice dimension is higher than 2. This is only a preamble to the rich phenomenology of diffusion in real-world networks which typically have complex topologies. Physical networks are notoriously heterogeneous and thus the graph dimension can differ from that of the embedding space \cite{Net_dimension}. Small-world structures like social and neuronal networks \cite{Small_world}, or self-similar structures such as intracellular networks \cite{Intracellular_transport}, are some examples of complex architectures. 

Besides the graph heterogeneity, the properties of diffusion can be affected by temporal heterogeneity in the waiting time before performing a step. Often, the presence of barriers hinders the diffusive passage and leads to sub-diffusive scaling of the standard deviation of the random walk displacement. An opposite effect is induced by mechanisms of active motion like motors that can increase the propagation speed leading to super-diffusive scaling of the standard deviation. 

Search efficiency is often crucial both for the survival of the random walker and the network itself, as is the case in biological systems, so optimization  of the search is paramount.  
What is the optimal search network architecture for each diffusive regime?
This question has recently motivated a considerable amount of work, starting with investigations of the effect of the location and connectivity of a specific target node on the global mean first passage time (GMFPT), defined as the MFPT averaged over an ensemble of starting nodes \cite{Global_MFPT}. More recently, for planar organelle networks, Ref.~\cite{Koslover1} studied the effect of loops on the overall network transport,  quantified by the GMFPT averaged over all the targets (TA-GMFPT). However, the graph structure of optimal search networks is still unknown.

\begin{figure}[h]
\centering
\includegraphics[width=8.5cm]{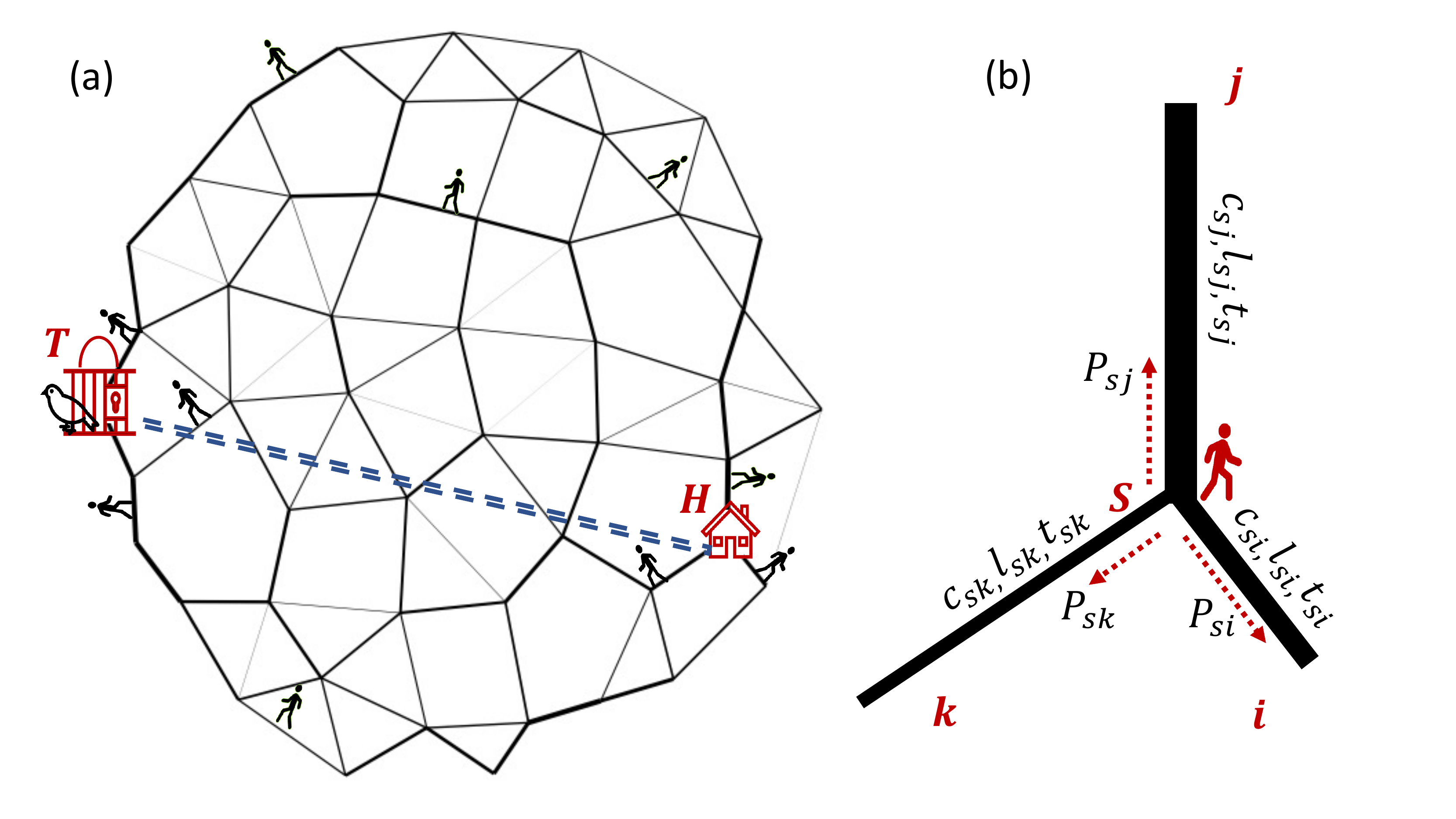}
\caption{(a) Sketch of a diffusive search for the target $T$, starting at a home node $H$. A shortcut (dotted path) affects both the MFPT to the target and the overall TA-GMFPT. (b) A junction on a weighted and spatially embedded graph. The conductivity $c_{si}$ and length $l_{si}$ of each link control the transition probabilities $P_{si}$ between neighboring nodes.}  \label{fig:cartoon}
\end{figure}

To optimize search efficiency, intuition suggests adding direct links to topologically distant pairs of nodes (Fig.\ref{fig:cartoon}). Any shortcut should reduce the MFPT time of the pair and provide an extra pathway to the rest. Counterintuitively, we find that the shortcut could also negatively affect the TA-GMFPT. In 1968,  Dietrich Braess demonstrated a paradoxical behavior of transport networks, where the addition of an extra road can worsen the overall traffic \cite{Braess_origin}. In our search networks, to avoid Braessian-like links, we propose a general optimization scheme under which each edge is adjusting its weight so that the TA-GMFPT or \textit{search time} of the graph is minimized.

In particular, we consider random walks on a set of spatially embedded, weighted graphs of $N$ nodes and $N_{e}$ edges. Two nodes $i$, $j$ can be linked with only one edge with weight $w_{ij}=\frac{c_{ij}}{l_{ij}}$. Note that the inverse weight can be thought of as the resistance $r_{ij} =w_{ij}^{-1}$ of the edge, in a resistor analogue of the random walk \cite{Snell}. Unlike previous work, see e.g. \cite{eigen_zhang}, here we consider both a length independent $c_{ij}$ which can be thought of as the conductivity of the edge, and the Euclidean distance between the two nodes $l_{ij}$.  The graph $G$ is either a natural representation of the system of interest (e.g. random walk of proteins) or constructed as a coarse-grained representation of a potential energy landscape. The motion of the particle can be treated as a Markov state model, with each Markov state corresponding to a node neighborhood. Transitions between neighborhoods are memoryless. We define the transition probability $P_{ij}$ according to which the random walker starting at node $i$ will jump to an adjacent node $j\in \mathcal{N}(i)$ as ${P}_{ij}=\frac{w_{ij}}{deg({i})}$, where $deg({i})\equiv \sum_{j\in \mathcal{N}(i)}w_{ij}$ is the weighted degree of node $i$ and $\mathcal{N}(i)$ is the set of nodes neighboring to node $i$.

In each link, the dynamics are treated as 1D anomalous diffusion with a mean transit time $t_{ij}$ through the link. This is the time when the standard deviation of the random walk displacement equals the link length $\sqrt{(\Delta x^2)}=l_{ij}$ and as such   $t_{ij}=\frac{l_{ij}^{2\beta}}{\mathcal{D}_{\beta}}$, where $\mathcal{D}_{\beta}$ is the diffusivity. The anomalous diffusion exponent $\beta$ captures two distinct regimes: $\beta<1$ for super-diffusive motion and $\beta>1$ for sub-diffusive \cite{diffusion_review}. Finally, we introduce the weighted waiting (or travel) time ${\tau}_{i}\equiv {\sum_{j\in \mathcal{N}(i)}w_{ij}t_{ij}}$, required for a random walker starting from node $i$ to transit to any of the neighbors $j\in \mathcal{N}(i)$. 

Let $T_{{i\rightarrow y}}$ be the MFPT from $i$ to target $y$. We calculate $T_{{i\rightarrow y}}$ in a recursive way employing the harmonic property, in which the MFPT from node $i$ to the target $y$ is the weighted average of the MFPTs of the neighbors $j$ of $i$,  plus the average waiting time to transit to any of the neighbors $T_{{i\rightarrow  y}}=\sum_{j\in \mathcal{N}(i),j\neq y}\frac{w_{ij}T_{j\rightarrow y}}{deg(i)}+\frac{\tau_{i}}{deg(i)}$. Generally, the recursive relation of the MFPT can be recovered from the scope of continuous time random walks as introduced by Montroll and Weiss \cite{Montroll}, using a spatiotemporal transition matrix to calculate the asymptotic behavior of the propagator in \cite{Grebenkov} (see Supplement).

We now consider the trapping problem of a random walker in the weighted graph $G$ when there is a set of target nodes $\{y\}$ which are perfect absorbers. The recursive relation for the MFPT can be expressed using the weighted Laplacian matrix $\boldsymbol{\mathcal{L}}$ where $\mathcal{L}_{ij}=w_{ij}$ for $i\ne j$ and $\mathcal{L}_{ii}=-\sum_{j\in \mathcal{N}(i)}w_{ij}$. If node $y$ is a target the $y^{th}$ rows and columns of the matrix $\boldsymbol{\mathcal{L}}$ have zero entries, and the $y^{th}$ row in the waiting time vector ${\boldsymbol{\tau}}$ is infinite and can be removed. Finally we obtain $T_{{i\rightarrow  y}}=\sum_{j, j\ne y}\mathcal{L}^{-1}_{ij} {\tau}_{j}$. To get a global estimate of the efficiency of the diffusive search for the graph $G$ we define the average pairwise MFPT over all pairs of nodes (TA-GFMPT) as $ \overline{\langle T\rangle}=\frac{1}{N(N-1)}\sum_{y} \sum_{i, i\ne y} T_{{i\rightarrow y}}$. 
If the Laplacian has a complete set of eigenvectors we can express $\boldsymbol{\mathcal{L}}^{-1}$ as a function of its non-zero eigenvalues and eigenvectors similarly to \cite{eigen_zhang}, (see Supplement), to obtain: 
\begin{equation}\label{important_result} \overline{\langle T\rangle}=\frac{1}{N-1}\tau_{G}\sum_{k=2}^{N}\frac{1}{\lambda_{k}}, 
\end{equation}
where $\lambda_1=0$ and $\lambda_k>0$ for $k>1$.

\begin{figure}[h]
\centering
\includegraphics[width=8.5cm]{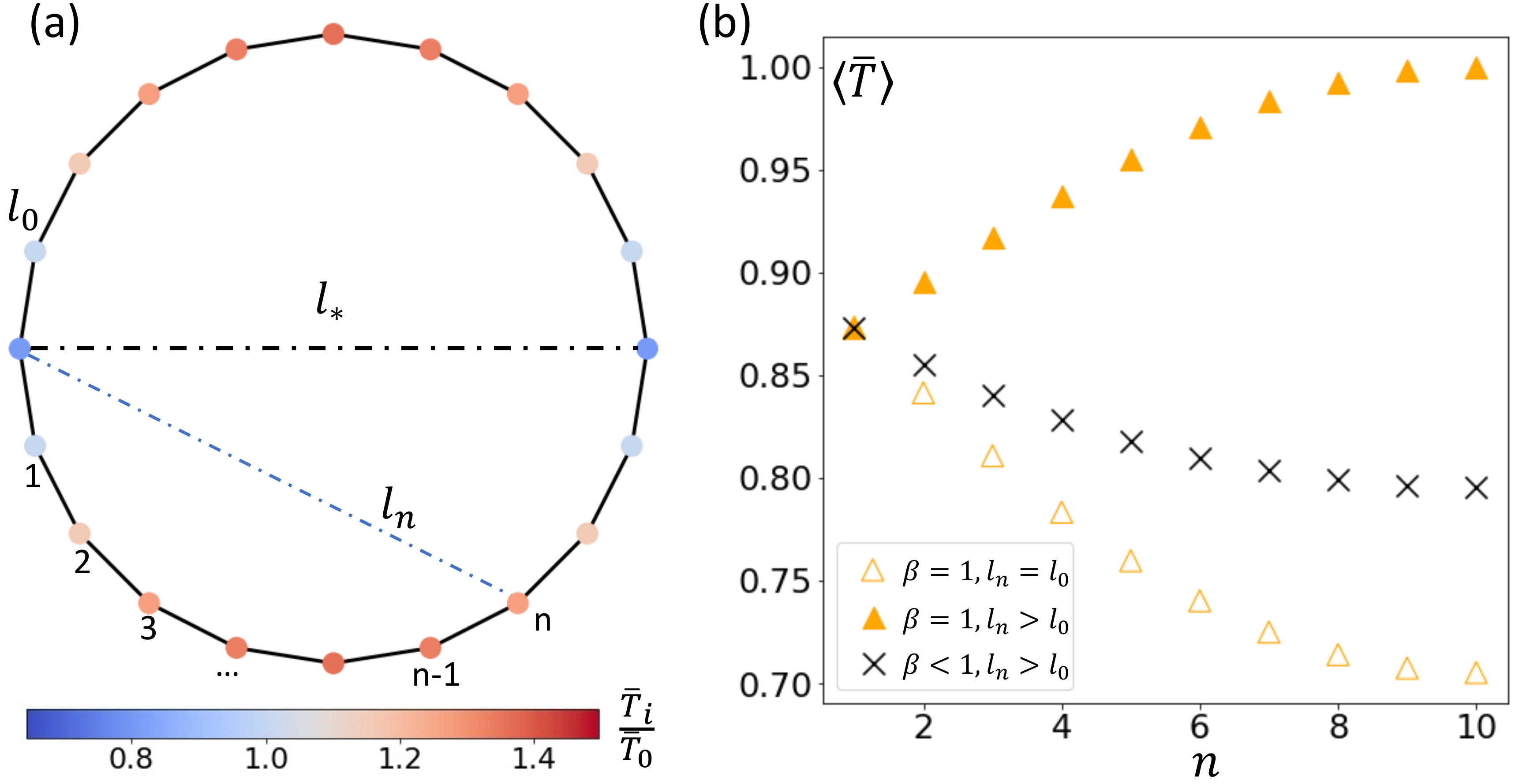}
\caption{(a) Diffusive search in an 1D periodic path, with lattice length $l_0$. The node labels correspond to the topological distance  $n$  from node $0$ and the corresponding shortcuts $s_n$ of geometrical length  $l_{n}$ are shown with dotted cords. The length of the longest shortcut $s_*$ is denoted as $l_{*}$. The color of the nodes signifies the GMFPT $\bar{T}_i$ for each target node once the longest shortcut is added. The short-circuited nodes have the lowest GMFPT. (b) Graph search time as a function of topological distance $n$, when one shortcut $s_n$ has been added to a ring of $N=20$ nodes. If all the links have identical lengths, adding shortcuts monotonically of increasing $n$ decreases the total search time for regular diffusion (hollow  triangles). When lengths are heterogeneous with $l_n>l_0$, and $\beta=1$, the addition of shortcuts monotonically increases the total search time (filled triangles), making the shortcuts Braessian edges. However, for super-diffusion, e.g. $\beta=0.5$, the shortcuts improve the total search time (crosses).}
\label{Braess_fig}
\end{figure}
The overall search time is related to the total waiting time $\tau_{G}=\sum_{i=1}^{N}\sum_{j\in \mathcal{N}(i)}\frac{1}{\mathcal{D}_{\beta}}w_{ij}l_{ij}^{2\beta}$ characterized by the anomalous diffusion exponent $\beta$. It is also related to the structural properties of the graph as expressed through the graph Laplacian eigenvalues $\lambda_{k}$ that are also related to the effective resistance of the graph $R_{G}=\frac{1}{N}\sum_{k=2}^{N}\frac{1}{\lambda_{k}}$. Here  $R_{G}=\sum_{i>j}R_{ij}$ and $R_{ij}$ is the effective resistance between nodes $i,\ j$ if a unit current is inserted on $i$ and retrieved at $j$ and the effective voltage drop is the MFPT \cite{Snell}. It is a metric that measures the strength of parallel paths between a pair of nodes and is less or equal to the smallest resistance path $L_{ij}\geq R_{ij}$. The equality holds for a network connected in series and becomes strict inequality once there is at least one parallel path. The relation between the effective resistance and the  TA-GMFPT is not surprising since the effective resistance is inversely proportional to the escape probability of the random walk \cite{Snell}. An infinite graph resistance implies the recurrence of the random walk and finite effective resistance implies transience.

We now examine the impact of topological modifications on the diffusive search efficiency of the graph. We begin by considering circular graphs $G(N,deg=2)$ that form a 1D periodic path with lattice constant $l_{0}$  (Fig.\ref{Braess_fig}.a). A shortcut $s_n$ with length $l_{n}$ is a direct link connecting two non-neighboring nodes with $n>1$ links separating them in the original lattice.  We make a distinction between two different cases: the non-spatially embedded graphs where all the edges are considered of equal length $l_{n}=l_{0}$ and the spatially embedded networks  where the length of an edge is given by the Euclidean distance between the connected nodes.

If all the edge lengths are equal then the transit time through each link is the same $t_{ij}= t_{0}$. This is analogous to  anomalous diffusion exponent $\beta=0$ which implies a link transit time independent of the length $t_{ij}=\frac{1}{\mathcal{D}_{0}}$. For equal lengths the graph mean first passage time  becomes $ \overline{\langle T\rangle}\approx W_{tot} {R}_{G}$ which is a convex function of the weights (or $c_{ij}$, as $l_{ij}=\mathrm{const}$) when the total weight is fixed $\sum_{i=1}^{N}\sum_{j\in \mathcal{N}(i)}w_{ij}=W_{tot}$ \cite{Minimize_resistance}. Consequently,  adding any link improves the search time and the optimal architecture is the complete graph \cite{Complete_optimal} (see Supplement).

However, in most physical networks there is heterogeneity in the length and conductivity distributions. The 1D spatially embedded ring $G$ reveals the competing relationship between topology, geometry, and diffusion. Suppose a shortcut $s_*$ with conductivity $c_{*}$ and length $l_{*}$ is added between antipodal nodes. This addition will have a twofold effect. The new weighted pathway will decrease the effective resistance $R_{G+s_{*}}<R_{G}$ maximally since it bridges the most distant nodes. On the other hand, it will increase the total waiting time $\tau_{G+s_{*}}=\tau_{G}+\frac{c_{*}l_{*}^{2\beta-1}}{\mathcal{D}_{\beta}}$.

Counterintuitively, we find that for regular and sub-diffusive exploration $\beta \geq 1$, the addition of any link longer than the lattice constant $l_*\geq l_{0}$ increases the total search time, although it increases the number of available paths in the network. This is an analog of the Braess' paradox, in the sense that increasing the available pathways can reduce the overall search efficiency. Such behavior is exemplified in various systems including traffic networks in which adding a road increases the total transportation time \cite{Anarchy_&_Braees}, or in power grid networks in which adding extra lines could reduce the performance and even promote blackouts \cite{Braess_power_grid}.\begin{figure}[ht]
    \centering
    \includegraphics[width=8.5cm]{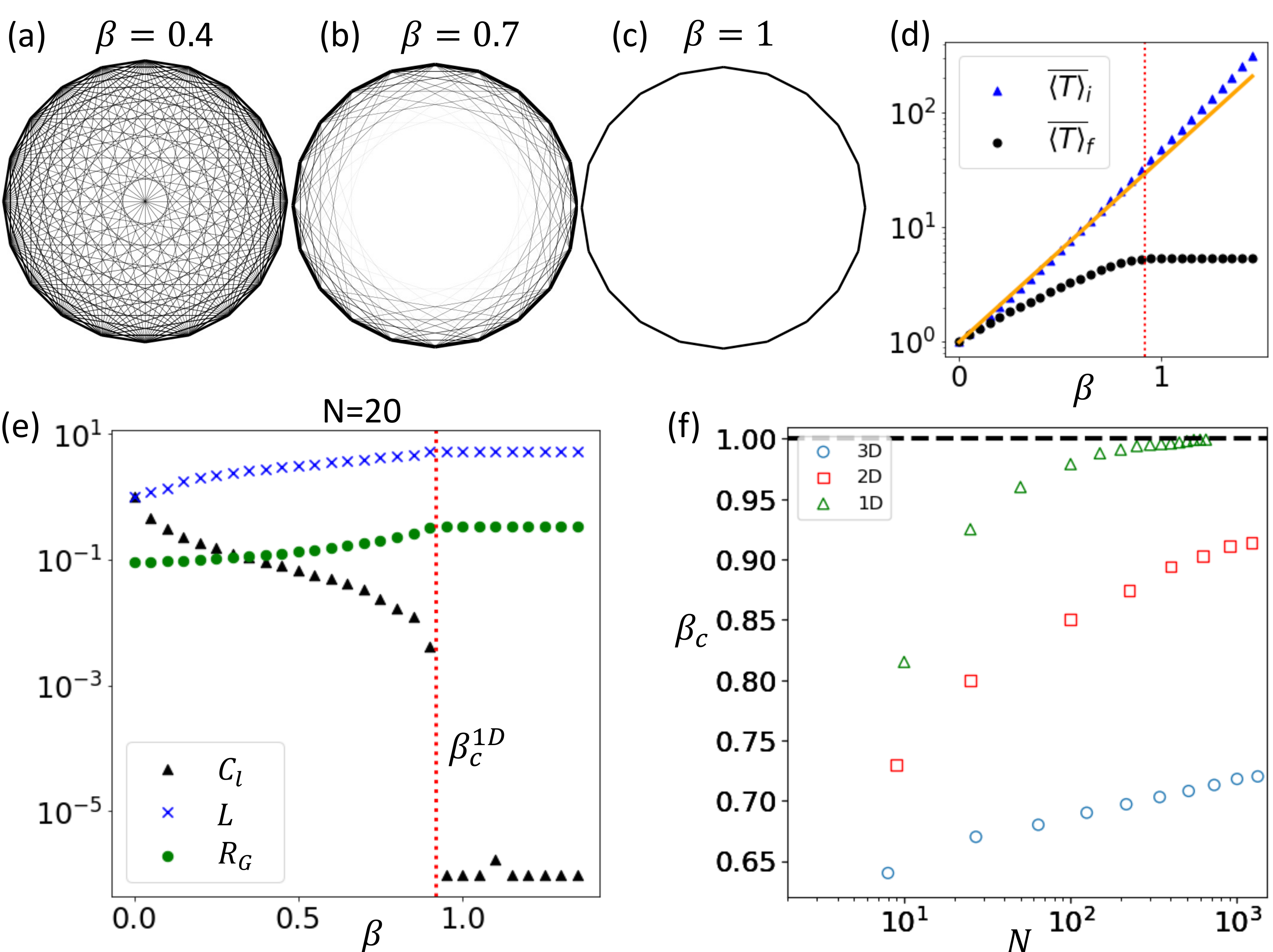}
    \caption{(a-c) Optimal search networks for different $\beta$. Panel (a) is a complete graph with nodes arranged on the perimeter of a circle (N=20). As $\beta$ increases, the shortcut's (chords) weight (width)   decreases while the weight of the short edges increases. For sub-diffusion $\beta \geq1$, only the shortest edges remain. (d) Initial, non-optimized search time for the all-to-all graph $\overline{\langle T\rangle}_i$ (blue triangles) and final, optimized search time $\overline{\langle T\rangle}_f$ (black circles) as a function of $\beta$. The orange line plots the exponential   $\overline{\langle T\rangle}_i\sim \langle l\rangle_{i}^{2\beta} $. (e) Both the mean topological shortest weighted path length $L$ (see supplement) and the graph effective resistance $R_{G}$ are monotonically increasing with $\beta$. The weighted clustering coefficient $C_l$ (black triangles) decreases  with $\beta$, and rapidly drops at $\beta_c$, highlighting the transition from dense graphs with long-range links  to sparse short-range lattices. (f) The crossover diffusion exponent  $\beta_{c}$, depends both on the size $N$ of the graph and the dimension of the embedding, where 1D, 2D, and 3D are shown in green, red, and blue respectively. }
    \label{complete_graph}
\end{figure}

Here we present a  systematic way to find the optimal graph, that minimizes the search time $\overline{\langle T\rangle}$ using gradient descent. At each iteration, the conductivity of each edge is allowed to adapt, while its length $l_{ij}$ is kept fixed. The updated conductivity at the $n+1$ iteration is $c^{n+1}_{ij}=c^{n}_{ij}+\delta c_{ij}$ and the weights $w^{n+1}_{ij}=\frac{c^{n+1}_{ij}}{l_{ij}}$. The variation of the conductivity $c_{ij}$ is given by the local rule $\delta c_{ij}\approx -v \frac{\partial E}{\partial c_{ij}}$ where $v$ is the step size, and E is the objective function 
\begin{equation}\label{adaptation_rule}
E=\overline{\langle T\rangle} +\lambda\left(\sum_{i=1}^{N}\sum_{j\in \mathcal{N}(i)}w_{ij}-W_{tot}\right)
\end{equation}
with $\lambda$ the Lagrange multiplier to ensure the conservation of the total weight $W_{tot}$ available to the system.

To investigate the impact of shortcuts on the search process, we  embed the nodes on a 1-D ring and connect all the possible node pairs (complete graph). The initial conductivity of each link is assigned equal to its length $c^{0}_{ij}=l_{ij}$, resulting in unitary initial weights $w^{0}_{ij}=1$. We then apply the adaptation rule to explore the optimized graph characteristics for different values of the diffusion exponent $\beta$. To analyze the network features, we employ robust topological measures, including the clustering coefficient ($C$) \cite{Clustering} and mean shortest path length ($L$) \cite{Strogatz}.

In 1D for super-diffusive motion ($\beta<1$) the optimal network structure permits shortcuts and is highly clustered,  see Fig.\ref{complete_graph}(a,b,e). For sub-diffusive motion ($\beta\geq 1$) long links are penalized and the optimal graph is the 1D ring-lattice, with minimal clustering coefficient and maximal shortest path length and effective resistance. The crossover at $\beta^{1D}_{c}\approx1$ is also captured by the search time. Fig.\ref{complete_graph}(d) shows the initial and optimal search times $\overline{\langle T\rangle}_{i,f}$ as a function of $\beta$. Note that the initial, non-optimized search time for the all-to-all graph $\overline{\langle T\rangle}_i$ (blue crosses) depends exponentially on the initial mean edge length $\langle l\rangle_i$, $\overline{\langle T\rangle}_i\sim \langle l\rangle_{i}^{2\beta} $(orange line). The mean edge length is defined as the weighted average of link lengths $\langle l \rangle \equiv \sum w_{ij} l_{ij}/W_{tot}$. The optimal TA-GMFPT $\overline{\langle T\rangle}_f$ (black dots) saturates for $\beta \geq1$ when the optimal graph becomes a ring lattice. For sub-diffusion, the optimal graph (ring) has   the minimal mean edge length  as $ \langle l\rangle = l_0$, and the search time scales as $\overline{\langle T\rangle}_f \sim l^{\beta}_{0}$ ($l_0$=1 in Fig.\ref{complete_graph}(d)).  

Due to finite size effects, the transition from dense to sparse graphs occurs asymptotically close to $\beta_{c}=1$ in 1D, as $N \rightarrow \infty$, see Fig.\ref{complete_graph}(f) (\emph{green}). The crossover exponent is a function of both the network size and the dimensionality.  As a final note, we observed the crossover at the same $\beta_{c}$ for any convex constraint function of the conductivities $c^{\gamma}_{ij}$, $\gamma\geq 1$, and even when the constraint was softened to an upper bound.

\begin{figure}[h]
    \centering
    \includegraphics[width=6cm]{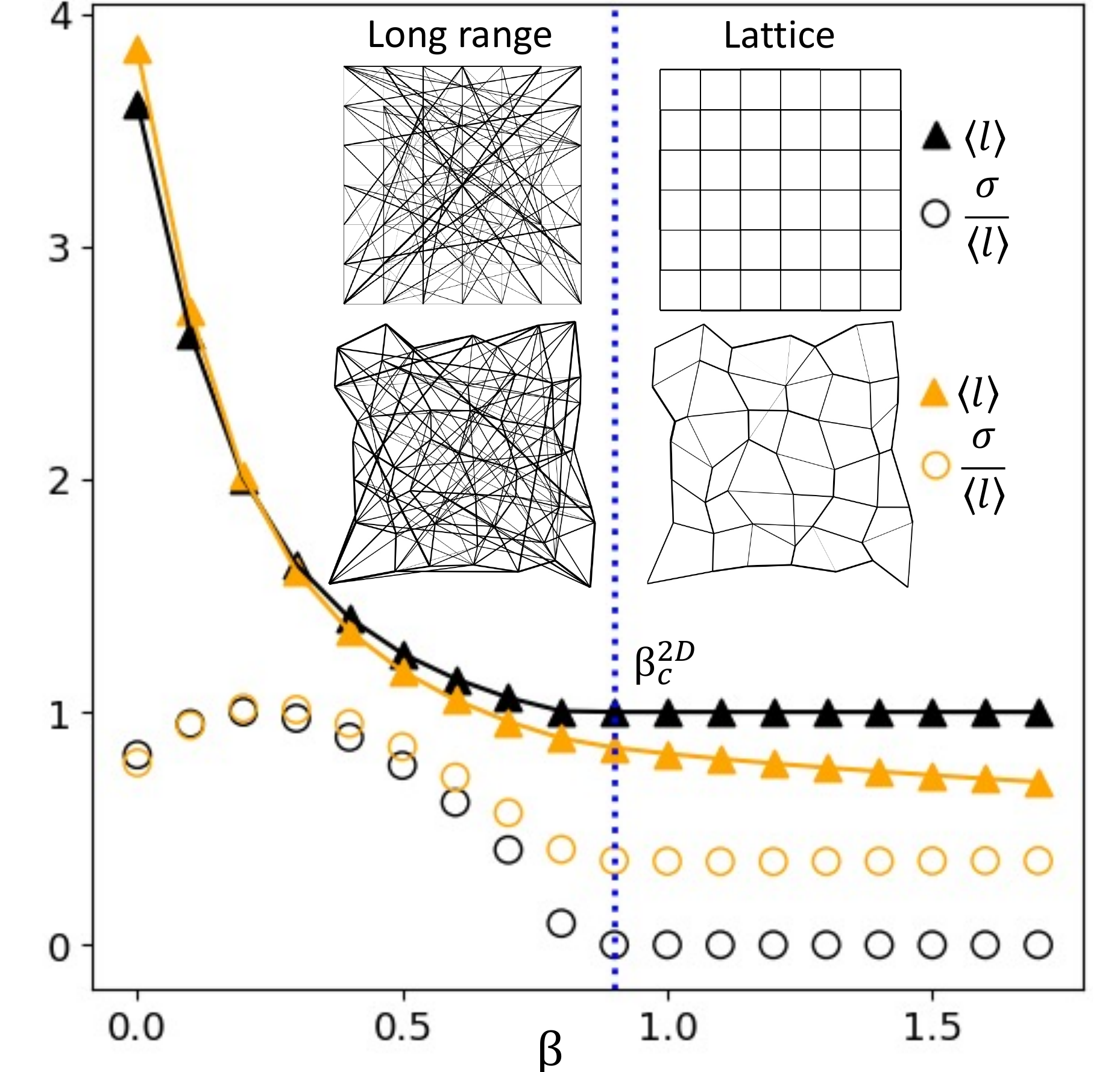}
    \caption{The mean link length $\langle l \rangle$ (triangles) and variation coefficient $\frac{\sigma}{\langle l \rangle}$ (circles) for the optimal search networks in square lattice and random node embeddings (black) and  (orange) respectively for N=144 nodes. For  $\beta<\beta^{2D}_{c}$, the networks  (left of the inset, smaller network size example) have long-range links. As $\beta$ increases to $\beta>\beta^{2D}_{c}$ only the shortest lengths survive. The networks with random node positions achieve a lower average edge length than the regular square and cubic lattices. In 2D the crossover occurs at $\beta^{2D}_{c}\approx 0.89$ (dotted line). In a 3D cubic lattice of $N=145$ nodes, $\beta^{3D}_{c}\approx 0.65$ (see Supplement).}
    \label{higher_dimensions}
\end{figure}

We now test how the optimal search  architecture depends on the dimension of the embedding space and the node positions. We study 2D and 3D using square and cubic lattice node embeddings respectively. In the regular lattice, we introduce  random links with probability $ p \sim 0.02 $ between unconnected pairs of nodes to obtain networks with high clustering and non-local connections as demonstrated in the examples of Fig.\ref{higher_dimensions} (inset).
Then we utilize a modified gradient descent algorithm \ref{adaptation_rule} to minimize the search time and we quantify the statistics of edge lengths of the optimal graphs. We find that, as the random walk propagation speed decreases ($\beta$ increases), both the average $\langle l \rangle $ and the variation coefficient $\sigma/\langle l \rangle$ of the edge length decreases in the optimal graphs. As Fig.\ref{higher_dimensions} shows, there is a transition from dense networks with long-range links to short-range sparse graphs with connections between geometrical nearest neighbors even in higher dimensions. The transition occurs at different anomalous diffusion exponents in different embedding dimensions  $\beta^{1D}_{c}>\beta^{2D}_{c}>\beta^{3D}_{c}$, as in Fig.~\ref{complete_graph}(f). When the nodes form a regular lattice, for $\beta>\beta_{c}$ the algorithm finds the minimal length graph which is a lattice with zero length variance $\frac{\sigma}{\langle l \rangle}=0$, shown in Fig.\ref{higher_dimensions} with black circles. 

Additionally, we examine the effect of random node positioning. We perturb the initial square or cubic lattice, $\boldsymbol{x^{r}_i}=\boldsymbol{x^{o}_i}+\boldsymbol{\delta x_i}$, where $\boldsymbol{x^{o}_i}$ is the regular lattice node position of node $i$ perturbed by $\boldsymbol{\delta x_i}$ whose components are drawn randomly from a uniform distribution $[-\frac{l_{0}}{2},\frac{l_{0}}{2}], $ where $l_{0}$ is the regular lattice spacing. Ultimately, we add random links with the same probability as for the regular graph case $ p \approx 0.02$. The network with random node positions achieves a lower average link length than the equivalent regular lattice, with the price of higher variance, as seen in Fig.\ref{higher_dimensions}. For completeness, the transition from a dense long-range graph to a sparse lattice can be captured in the clustering coefficient as presented in Fig.\ref{complete_graph}(e) (see Supplement).


To shed light on the crossover from dense to sparse optimal search graphs, we investigate the conditions under which a link addition can reduce the overall search time. Assume $G$ is a graph with $N$ nodes arranged approximately on a $D$ dimensional hypercubic lattice with lattice spacing $l_0\simeq 1$ and conductivity $c_0=1$ resulting in  $w_{0}=r_0=1$. The graph $G+s_{*}$ is an extended graph with an additional shortcut $s_{*}$ of length $l_{*}\approx N^{\frac{1}{D}}$ and conductivity $c_{*}=1$ resulting in $w_{*}=\frac{1}{l_*}$. From Eq.~\ref{important_result}, we see that the graph's search time  is $\overline{\langle T\rangle}_{G} = \frac{N}{N-1} \tau_{G} R_{G}$ and is improved after the shortcut addition if:  \begin{equation}\label{shortcut_condition}\frac{\overline{\langle T\rangle}_{G+s_{*}}}{\overline{\langle T\rangle}_{G}}=\frac{\tau_{G+s_*}}{\tau_{G}} \frac{R_{G+s_{*}}}{R_{G}}<1.
\end{equation}
The addition of $s_{*}$ always increases the total edge length and consequently, the total travel time $\tau_{G+s_{*}}=\tau_{G}+ \frac{2}{\mathcal{D_{\beta}}}l^{2\beta -1}_{*}$. If $G$ is a regular lattice with $N_e$ unit length edges it has total transit time $\tau_{G}=2 \frac{1}{\mathcal{D_{\beta}}}N_{e} $ which implies:
\begin{equation}\label{time_ratios}
\frac{\tau_{G+s_*}}{\tau_{G}} =1+ \frac{N^{\frac{2\beta-1}{D}}}{{N}_{e}} >1.
\end{equation}

Conversely, $s_{*}$ provides 
 additional pathways and reduces the graph's resistance $R_{G+s_{*}} < R_{G}$ (see Supplement). If $(i,j)$ is the pair of nodes with the maximum effective resistance $R_{ij}^{max}$ on the lattice then: \begin{equation}\label{graph_res_bound}
  1 > \frac{{R}_{G+s_{*}}}{R_{G}}> \left(\frac{R_{ij}^{max}}{r_{*}}+1\right)^{-1}.
\end{equation} 
If the shortcut's weight is very small,  its resistance $r_*$ is very large  $r_{*}\gg R^{max}_{ij}$, then $\frac{R_{G+s_{*}}}{R_{G}}\rightarrow 1$ and the shortcut does not improve the search time. It is noteworthy that the $\frac{R_{G+s_{*}}}{R_{G}}$ scaling depends on the lattice dimension. 

\begin{figure}[h]
    \centering
    \includegraphics[width=8.6cm]{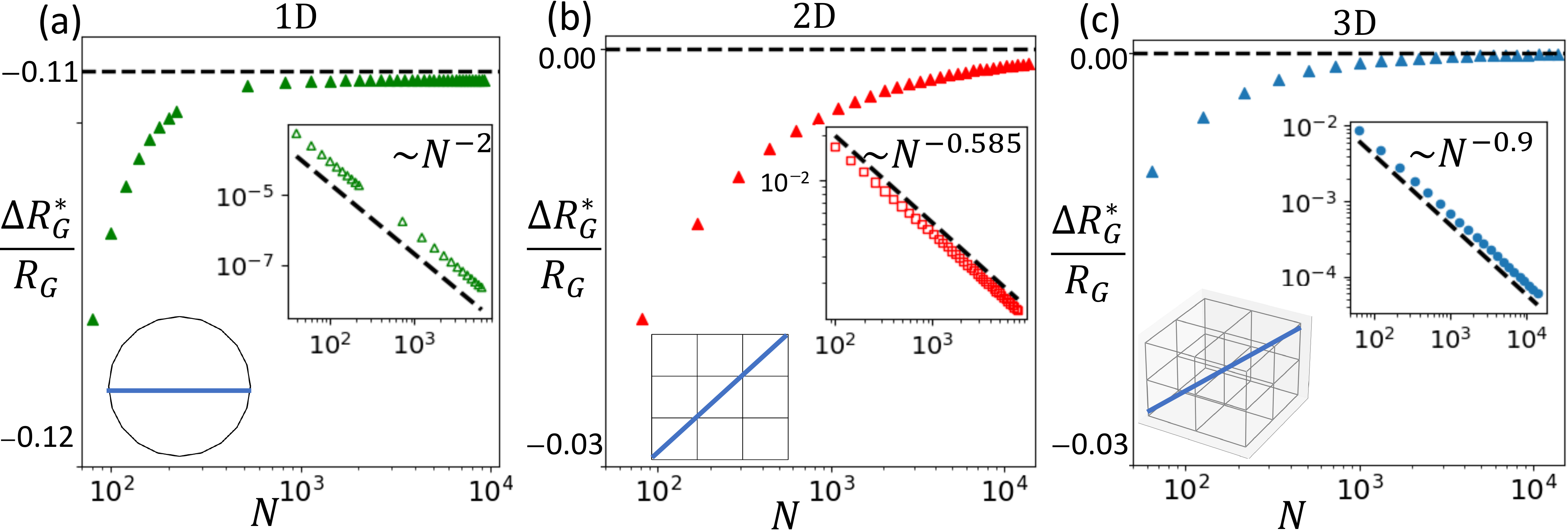}\caption{Graph resistance reduction $\frac{\Delta R^{*}_{G}}{R_{G}}\equiv \frac{{R}_{G+s_{*}}-R_{G}}{R_{G}}$ after the addition of a shortcut  (blue link on inset graphs) as a function of nodes $N$. (a) In 1D the addition of shortcuts always improves the resistance, as $\frac{\Delta R^{*}_{G}}{R_{G}}<0$. In (b) 2D and (c) 3D for $N\rightarrow \infty$  there is no improvement by the shortcut addition $\frac{\Delta R^{*}_{G}}{R_{G}}\rightarrow 0$. The logarithmic inset plots reveal the power law scaling of $\frac{\Delta R^{*}_{G}}{R_{G}}$ with the system size (dashed lines). The exponents were obtained through a linear  fitting.}
    \label{shortcut_and_eff_res}
\end{figure}

For a 1D ring embedding, the effective resistance between antipodal nodes scales as $R^{max}_{ij} \sim N $ and the shortcut resistance scales as $r_{*} \sim {N}$. As Fig.\ref{shortcut_and_eff_res}\emph{a} shows, $\frac{R_{G+s_{*}}}{R_{G}}<1$  and it is enough to ensure that $\frac{\tau_{G+s_*}}{\tau_{G}}$ is finite, to satisfy Eq.~\ref{shortcut_condition} which implies $\beta^{1D}_{c} \leq 1$. 

In a 2D square lattice, the maximum effective resistance scales as $R^{max}_{ij}\sim lnN $ for $N\gg 1$ \cite{Resistance_lattice} and the maximum shortcut length is  $l_{*}\sim \sqrt{2 N } $ with resistance $r_{*}=\sqrt{2N}$. Asymptotically,  $r_{*}\gg R^{max}_{ij}$ and the shortcut does not reduce the search time $\frac{{R}_{G+s_{*}}}{R_{G}}\rightarrow 1$, Fig.\ref{shortcut_and_eff_res}\emph{b}. In particular, $\frac{{R}_{G+s_{*}}}{R_{G}}\sim 1-N^{-0.585}$ and  $\frac{\tau_{G+s_*}}{\tau_{G}}\sim 1+N^{\beta-\frac{3}{2}} $ and \ref{shortcut_condition} implies that $\beta^{2D}_{c}<0.915$ (obtained through line fitting of insets in Fig.~\ref{shortcut_and_eff_res}\emph{b}). 

In 3D the random walk is transient and the effective resistance between any pair of nodes is finite regardless of their distance $R_{ij}^{max}=const$ \cite{Resistance_lattice}. The maximum shortcut length is $l_{*} \sim  N^{\frac{1}{3}}$ with resistance $r_{*}\sim N^{1/3}$ which for large enough network gives $r_{*}\gg R^{max}_{ij}$. As seen in  Fig.\ref{shortcut_and_eff_res}\emph{c}, $\frac{{R}_{G+s_{*}}}{R_{G}}\sim 1-N^{-0.9}$ and  $\frac{\tau_{G+s_*}}{\tau_{G}}\sim 1+N^{\frac{\beta-4}{3}} $ then  \ref{shortcut_condition} implies that $\beta^{3D}_{c}<0.65$ (line fitting). 

Previous work has characterized diffusive search as compact if $\beta<d_{f}$ for the random walk dimension $\beta$ and the fractal dimension $d_{f}$ of the graph,  or non-compact if $\beta> d_{f}$ \cite{Geometry_benichou}. The marginal case is when $\beta=d_{f}$.
In our analysis, we find that compactness is not enough to capture the crossover of the optimal architecture from long-range connections to a short-range lattice architecture in continuous time random walks. The transition depends on the type of diffusion, along with the graph dimension and the embedding dimension. In 1D we obtain, analytically and numerically through optimization, that for sub-diffusion $\beta \geq \beta^{1D}_{c}=1 $ optimal search favors short-range links, while for super-diffusion $\beta < 1 $ non-local links increase the search efficiency. We note that the random walk is non-compact when $\beta>0.5$ in 1D. In 2D the crossover occurs for $\beta^{2D}_{c}< 1$ and the condition for compactness is $\beta>1$. Finally, in 3D compactness occurs for $\beta<1.5$ but the optimization indicates that the crossover occurs for lower values, as $\beta^{3D}_{c}\approx 0.71$ is the highest we obtained for $N=12^{3}$ Fig.\ref{complete_graph}(\emph{f}). 

Over the last two decades, the interest in first encounter properties keeps growing driven by the new experimental discoveries that showcased diffusion as a ubiquitous mechanism of transport and exploration in complex environments. The notion of length and travel time in diffusive search has proven crucial for the modeling of these processes, as physical networks are spatially embedded with links that often span various length scales, e.g., in chemical transcription networks \cite{Geometry_benichou} or intracellular transport networks \cite{Experiment_ER}. In particular, the transport of intracellular components relies on a variety of active and passive mechanisms, ranging from the diffusive spreading of small molecules over short distances to motor-driven motion across long distances. This points to the principle that biology has already developed mechanisms to deal with spatial heterogeneity \cite{Intracellular_transport}.

In this letter, we conducted a systematic investigation of the topological and geometrical features that a weighted and spatially heterogeneous network should have, to achieve minimal mean first passage times between any pair of nodes. A first realization is that once the transit time through an edge depends on the edge's Euclidean length, the travel time is characterized by the standard deviation of the random walk, which sets a time scale for propagation. As the epitome of this interplay, we show that for any sub-diffusive exponent $\beta\geq 1$, the addition of any extra link with a length greater than the average lattice deteriorates the overall transport time, even though it can bridge topologically distant nodes. This Braessian behavior is a consequence of the super-linear increase in the waiting time that cannot be compensated by the reduction in the effective resistance. 

Finally, we propose an adaptation rule according to which, each link updates its conductivity so that it reduces the overall transport time while conserving the available material resources. The optimization algorithm reveals a crossover of the network structure as a function of the diffusivity exponent $\beta$ and the embedding dimension. For $\beta < \beta^{D}_{c}$ we obtain dense graphs with long-range shortcuts, while for $\beta \geq \beta^{D}_{c}$ the optimal networks have a short-range architecture with connections only between geometrical nearest neighbors. We believe that this optimization approach might give insights into the potential mechanisms that highly optimized biological systems employ to solve the problem of efficient exploration in various length scales. \begin{acknowledgments}
The authors would like to thank A. Winn and J. Khoury for their useful comments on this manuscript and  S. Wong for discussions. This research was supported by NSF Grant No. PHY-1554887,  the  University  of  Pennsylvania  Materials Research Science and Engineering Center (MRSEC) through Grant No. DMR-1720530,
and the Simons Foundation through Grant No. 568888.\end{acknowledgments}


\bibliography{apssamp}

\end{document}